\documentclass[12pt]{article}
\catcode`\@=11
\@addtoreset{equation}{section}

\global\arraycolsep=1pt
\def\fnote#1#2{\begingroup\def\thefootnote{#1}\footnote{#2}\addtocounter
{footnote}{-1}\endgroup}

\usepackage{amsbsy,amssymb,latexsym,amsfonts}
\usepackage{mathrsfs}

\RequirePackage[dvips,usenames]{color}
\definecolor{fireblick}{rgb}{0.698039,0.133333,0.133333}

\newcommand{\beq}{\begin{equation}}
\newcommand{\eeq}{\end{equation}}
\newcommand{\bea}{\begin{eqnarray}}
\newcommand{\eea}{\end{eqnarray}}

\newcommand{\bC}{{\mathbb C}}

\newcommand{\bZ}{{\mathbb Z}}

\newcommand{\CP}{{\mathcal P}}
\newcommand{\CN}{{\mathcal N}}

\newcommand{\CD}{{\mathcal D}}
\newcommand\SD{\mathscr{D}}

\newcommand{\CF}{{\mathcal F}}

\newcommand\CQ{\mathscr{Q}}

\newcommand\BN{{\textbf{\textrm{N}}}}%{\boldsymbol{N}}

\renewcommand\Im{{\mathrm{Im}}}
\renewcommand\Re{{\mathrm{Re}}}

\newcommand\intd{{\int\!d}{}}

\newcommand\bi{{\bar{i}}}
\newcommand\bj{{\bar{j}}}
\newcommand\bk{{\bar {k}}}
\newcommand\ui{{\underline{i}}}
\newcommand\uj{{\underline{j}}}
\def\Tr{\mathop{\rm Tr}}

\newcommand\xiD{\xi_D}
\newcommand\xiDb{\bar{\xi}_D}
\newcommand\BD{{\boldsymbol{D}}}

\newcommand\diag{\mathrm{diag}}

\begin{document}

%\hfill
%\filename
%\\
\hfill
September, 2007
\\
\hfill
OCU-PHYS 275
\\
\hfill
OIQP-07-10

\bigskip

\begin{center}
{\bf\Large
$\CN=2$ Quiver Gauge Model\\
and\\
Partial Supersymmetry Breaking
}

\bigskip\bigskip
{
H. Itoyama$^{a,b}$\fnote{$\dag$}{e-mail: \texttt{itoyama@sci.osaka-cu.ac.jp}}
\quad, \quad
K. Maruyoshi$^a$\fnote{$*$}{e-mail: \texttt{maruchan@sci.osaka-cu.ac.jp}}
\quad and \quad
M. Sakaguchi$^c$\fnote{$\ddagger$}{e-mail: \texttt{makoto\_sakaguchi@pref.okayama.jp}}
}
\end{center}

\bigskip

\begin{center}
$^a$ \it Department of Mathematics and Physics,
Graduate School of Science\\
Osaka City University\\
\medskip

$^b$ \it Osaka City University Advanced Mathematical Institute
(OCAMI)\\

\bigskip

3-3-138, Sugimoto, Sumiyoshi-ku, Osaka, 558-8585, Japan\\

\bigskip

$^c$ {\it Okayama Institute for Quantum Physics\\
1-9-1 Kyoyama, Okayama 700-0015, Japan}

\end{center}

\vfill

\begin{abstract}
  We construct an action of $\CN=2$ affine $A_n$ quiver gauge model having non-canonical kinetic terms and 
  equipped with electric and magnetic FI terms.
  $\CN=2$ supersymmetry is shown to be broken to $\CN=1$ spontaneously
  and $\CN=1$ multiplets realized on the vacua are given.
  We also mention the models with different gauge groups.
  It is argued that
  the affine $A_1$ quiver gauge model provides a dynamical realization to approach 
  the Klebanov-Witten $\CN=1$ fixed point.
\end{abstract}

\newpage
%%%%%%%%%%%%%%%%%%%%%%%%%%%%%%%%%%%%%%%%%%%%%%%%%%%%%%%%%%%%
\section{Introduction}
  Supersymmetry has become one of the most remarkable and attractive ideas in theoretical physics.
  In particular, various investigations beginning with \cite{Seiberg, SW}
  have been made on $\CN=2$ supersymmetric Yang-Mills theory in four dimensions, 
  taking advantage of its powerful properties.
  Furthermore, we can extract the important information of $\CN=1$ super Yang-Mills theory, 
  such as the low energy effective superpotential, 
  breaking $\CN=2$ supersymmetry to $\CN=1$ by a superpotential \cite{SW, de Boer, Dijkgraaf:2002fc}.
  
  On the other hand, in view of the fact
  that superstring theories produce, in some backgrounds, extended supersymmetry in four dimensions 
  and have no adjustable parameter, 
  it is natural to consider spontaneous breaking of the extended supersymmetry
  so as to obtain more realistic $\CN=1$ supersymmetric models.
  Although it had been argued that, based on the supercharge algebra, 
  rigid $\CN=2$ supersymmetry is not spontaneously broken to $\CN=1$,
  a loophole has been first pointed out in \cite{Hughes:1986dn}
  by the argument based instead on the supercurrent algebra 
  which has been modified by an additional space-time independent term.
  In \cite{APT} and \cite{FIS1, FIS2, FIS3}, $\CN=2$, $U(1)$ and $U(N)$ gauge models 
  with $\CN=2$ vector multiplet only have been constructed, 
  establishing this modification of the algebra by introducing magnetic Fayet-Iliopoulos (FI) term.
  It was shown that the partial breaking of $\CN=2$ supersymmetry indeed occurs in such models.
  (See also \cite{KMG, Fujiwara, IM2} for related discussions and \cite{sugra} for supergravity.)
  
  In the $U(N)$ gauge theory which contains only the $\CN=2$ vector multiplet, 
  the magnetic FI term which causes the partial breaking can be easily introduced
  in the harmonic superspace formalism \cite{HS}(see \cite{HSbook} for a review)
  as a constant shift of the auxiliary field \cite{FIS3, FIS4}.
  In addition, it was shown that partial supersymmetry breaking can occur
  even in the presence of hypermultiplets in the adjoint representation.
  However, the addition of hypermultiplets in fundamental representation makes it difficult, 
  as pointed out in \cite{PP, Ivanov}.
  In this paper we overcome this difficulty by considering 
  a model with hypermultiplets in bi-fundamental representation, 
  whose matter content is described by a quiver diagram.
  This model is, therefore, a quiver gauge model. 
  We will show that, in addition to electric FI term,
  it is possible to introduce a magnetic FI term for any $\CN=2$ quiver gauge theory.
  This statement leads to the conclusion that in generic $\CN=2$ quiver gauge theory with these terms, 
  $\CN=2$ supersymmetry can be broken to $\CN=1$ spontaneously.
  As an illustration, we will describe this explicitly in a specific model, affine $A_1$ quiver gauge model, 
  focusing on the Coulomb branch. 
  
  This model may seem reminiscent of the one discussed in \cite{Klebanov:1998hh}: 
  a flow, by a mass deformation,
  from the $\CN=2$ affine $A_1$ theory on the world volume of the D3-branes at $\bC^2/\bZ_2$ orbifold singularity 
  to $\CN=1$ quiver gauge theory on that at conifold singularity.
  Indeed, we can show that in special points of the Coulomb branch 
  the mass spectrum is the same as that of the theory at the conifold singularity.
  The remarkable point of our model is that the masses are produced dynamically,
  and thus we can dynamically approach the theory on conifold, namely conifold geometry.

  The organization of this paper is as follows.
  In section 2, we construct $\CN=2$ affine $A_{n-1}$ quiver gauge model equipped with electric and magnetic FI terms.
  The necessary condition to introduce the magnetic FI term 
  without breaking $\CN=2$ supersymmetry in the action is examined.
  We also mention the cases of different types of quiver gauge theories. 
  As an illustration, we mainly consider one of the simplest models, affine $A_1$ quiver gauge model,
  in the subsequent sections.
  In section 3, we derive the scalar potential which is needed in the analysis of the vacua.
  We show, in section 4, that $\CN=2$ supersymmetry is broken to $\CN=1$ spontaneously on the Coulomb branch
  by observing the mass spectrum and the appearance of the Nambu-Goldstone fermion.
  As an application of our model, in section 5, 
  we consider the dynamical realization of $\CN=1$ quiver gauge theory on the world volume of the D3-branes 
  at the conifold singularity which has been considered in \cite{Klebanov:1998hh}.
  The notations on the harmonic superspace used in this paper are collected in appendix.

%%%%%%%%%%%%%%%%%%%%%%%%%%%%%%%%%%%%%%%%%%%%%%%%%%%%%%%%%%%%
\section{$\CN=2$ quiver gauge model}
  When we place $N$ D3-branes at $A_{n-1}$ orbifold $\bC^2/\bZ_n$, 
  the gauge theory realized on the world volume is $\CN=2$ affine $A_{n-1}$ quiver gauge theory \cite{DM}, 
  namely $U(N)^n$ gauge theory composed of 
  $n$ vector multiples $V^{++}_I$ in adjoint representation
  and $n$ hypermultiplets $q^+_I$ in  bi-fundamental representation, where $I=1,\cdots,n$.
  In this section, we will construct $\CN=2$ affine $A_{n-1}$ quiver gauge model, 
  which has the same matter content as above, 
  but with non-canonical kinetic terms and electric and magnetic FI terms.
  We examine a necessary condition to introduce the magnetic FI term 
  without breaking $\CN=2$ supersymmetry in the action.
  This magnetic FI term causes the partial spontaneous 
  breaking of $\CN=2$ supersymmetry as will be seen in the next section.
  
  In the case with additional D5-branes wrapping on non-trivial $S^2$s, as in \cite{DM, Klebanov:2000hb}, 
  the gauge group of the world volume theory is $\prod_i U(N_i)$ 
  and the rank of gauge group of each node is in general different.
  Although we will not consider the gauge model with this matter content explicitly, 
  we will mention the condition necessary to introduce
  the magnetic FI term at the end of this section.
  
  The action for the vector multiplet of the $U(N)^n$ gauge symmetry is given by
    \begin{eqnarray}
    S_V
    &=&  - \frac{i}{4}\int d^4x  \sum_{I=1}^n [(D)^4 \CF_I(W_I)
         - (\bar D)^4 \bar\CF_I(\bar W_I)]~
           \label{SV}
    \end{eqnarray}
  where $W_I$ is the curvature of $V^{++}_I$ and $(D)^4=\frac{1}{16}(D^+)^2(D^-)^2$.
  See appendix for the concrete form.
  $\CF_I$ is the prepotential for the $I$-th $U(N)$ which is denoted by $U(N_I$)
    \footnote{From now on we use this notation. 
              But keep in mind that we are considering $U(N)^n$ gauge model with the
               same ranks
              }.
  $V^{++}_I$ is composed of a complex scalar $\phi_I$, $SU(2)$ doublet Weyl fermions $\lambda_I^A$
  ($A=1,2$ labels $SU(2)$ automorphism of $\CN=2$ supersymmetry) and a real auxiliary field $D_I^{(AB)}$,
  which transform as adjoint representation of $U(N_I)$.
  The $U(N_I)$ gauge group is generated by $t_a^I$ ($a=0,1,\cdots,N_I^2-1$),
  and $t_0^I$ represents the overall $U(1)$ part of $U(N_I)$.
  As was done in \cite{FIS3}, (\ref{SV}) reduces in components to
    \begin{eqnarray}
    S_V
    &=&    \int d^4x\sum_{I=1}^n \Bigg[
           %% gauged kinetic term
         - \Im \CF_{I|ab}\CD^m\phi_I^a \CD_m\bar\phi_I^b
         - \frac{1}{2}\bar{\CF}_{I|ab}\bar\lambda_I^{aA}\bar\sigma^m\CD_m\lambda_{IA}^b
         + \frac{1}{2}{\CF}_{I|ab}\lambda_I^{aA}\sigma^m\CD_m\bar\lambda_{IA}^{b}
           \nonumber \\ 
    & &  + \frac{1}{4}\Im\CF_{I|ab}D_I^{a\,AB}D_{IAB}^b
         + \frac{i}{4}
           \CF_{I|abc}
           \lambda_I^{aA} \lambda_I^{bB}
           D_{IAB}^c
         - \frac{i}{4}\bar{\CF}_{I|abc}
           \bar\lambda_I^{aA}\bar\lambda_I^{bB}
           D_{IAB}^c
           \nonumber \\
    & &  - \frac{1}{4}\Im \CF_{I|ab}v_{Imn}^av^{b\,mn}_I
         - \frac{1}{8}\Re\CF_{I|ab}\varepsilon^{mnpq}v^a_{Imn}v^b_{Ipq}
           \nonumber \\
    & &  - \frac{i}{4}(
           \CF_{I|abc}\lambda_I^{aA}\sigma^{mn}\lambda^{b}_{IA} 
         - \bar{\CF}_{I|abc}\bar\lambda_I^{aA}\bar\sigma^{mn}\bar\lambda^b_{IA} 
           )v^c_{Imn}
           \nonumber \\
    & &  - \frac{i}{12}\CF_{I|abcd}(\lambda^{aA}_I\lambda^{bB}_I)(\lambda^c_{IA}\lambda^d_{IB})
         + \frac{i}{12}\bar{\CF}_{I|abcd}
           (\bar\lambda^{aA}_I\bar\lambda^{bB}_I)(\bar\lambda^c_{IA}\bar\lambda^d_{IB})
           \nonumber \\
    & &  + \frac{1}{2}\Im\CF_{I|ab}
           \left[
           \bar\lambda^{aA}_If^b_{cd}(i\sqrt{2}\phi_I^c)\bar\lambda^d_{IA}
         + %\frac{1}{2}\Im\CF_{ab}
           \lambda^{aA}_If^b_{cd}(-i\sqrt{2}\bar\phi_I^c)\lambda^d_{IA}
           \right]
%\nonumber\\&&
%-\frac{\sqrt{2}}{4}{\CF}_{ab}
%\psi^{ai}f^b_{cd}\bar\phi^c\psi^d_i
%-\frac{\sqrt{2}}{4}\overline{\CF}_{ab}
%\bar\psi^{ai}f^b_{cd}\phi^c\bar\psi^d_i
%+\frac{i}{4}\CF_a
%f^a_{bc}\bar\psi^{bi}\bar\psi^c_i
%-\frac{i}{4}\overline{\CF}_a
%f^a_{bc}\psi^{bi}\psi^c_i
           \nonumber \\ 
    & &  + {\frac{1}{2}}\Im\CF_{I|ab}\,f^a_{cd}\bar\phi^c_I\phi^d_I\,
           f^b_{ef}\bar\phi^e_I\phi^f_I
%~~~~
%\sim~~g^{ab}\fD_a\fD_b
%
%+\frac{\sqrt{2}}{4}\CF_a
%f^a_{be}f^e_{cd}\bar\phi^b\bar\phi^c\phi^d
%+\frac{\sqrt{2}}{4}\overline{\CF}_a
%f^a_{be}f^e_{cd}\phi^b\phi^c\bar\phi^dh
           ~~\Bigg]~.
           \label{SV:component}
    \end{eqnarray}
  where the symbol $\CF_{I|ab\dots}$ denotes the derivative of $\CF_I$ with respect to $\phi_I^a$, $\phi_I^b, \dots$.

  Let us introduce the $\CN=2$ hypermultiplets $q^+_I{}^i{}_\bj$ 
  ($i = 1, \ldots, N_I$, $\bi = \bar{1}, \ldots, \bar{N}_{I+1}$)
  transforming as ($\BN_I,\bar{\BN}_{I+1}$) under $U(N_I)\times U(N_{I+1})$: 
    \begin{eqnarray*}
    \begin{array}{cccccc}
          & U(N_1)   ~~&U(N_2)    ~~&U(N_3)    ~~&\cdots    ~~&U(N_n)    \\
    q^+_1 & \BN_1      &\bar{\BN}_2    &1    &\cdots    &1    \\
    q^+_2 & 1          &\BN_2    &\bar{\BN}_3    &\cdots     &1    \\
    \vdots&    &    &    &    &    \\
    q^+_n & \bar{\BN}_1   & 1   &1    &    &\BN_n ~.   \\
    \end{array}
    \end{eqnarray*}
%  (As the hypermultiplet $q^+_I$ is composed of two of $\CN=1$ chiral multiplets
%  $(A_I,\bar B_I)$, $A_I$ and $B_I$ transform as $(\BN_I,\bar{\BN}_{I+1})$
%  and $(\bar{\BN}_I,\BN_{I+1})$, respectively.)
  The matter part of the action is
    \bea
    S_q
     =   - \int d u d \zeta^{(-4)} 
           \sum_{I=1}^n
           (\tilde{q}_I^{+})^\bi{}_{i} (\SD^{++} q^+_I)^i{}_{\bi}~
           \label{Sq}
    \eea
  where $d\zeta^{(-4)}=d^4y d^4\theta^+$  and
    \bea
    (\SD^{++} q^+_I)^i{}_\bi
    &=&     D^{++} q^+_I{}^i{}_\bi 
          + i V_I^{++ a} (t_a^I)^i{}_{j} q^+_I{}^j{}_\bi
          - i V_{I+1}^{++ a} (\bar{t}_a^{I+1})^{\bj}{}_{\bi} q^+_I{}^i{}_\bj ~.
            \label{Dq}
    \eea
  As explained in appendix, after eliminating infinitely many auxiliary fields,
  a hypermultiplet $q^+_I$ contains $SU(2)$ doublet complex scalars $Q_{IA}$,
  and a pair of Weyl fermions, $\psi_I$ and $\bar{\kappa_I}$.
  In components, the matter action $S_{q}$ reduces to
    \begin{eqnarray}
    S_{q}
    &=&    \int d^4 x \sum_{I=1}^n\Big[
         - \bar Q^A_I{}^\bi{}_i\CD^m\CD_m Q_{IA}{}^i{}_\bi
         - \frac{i}{2}\bar\psi_I{}^\bi{}_i\bar\sigma^m\CD_m\psi_I{}^i{}_\bi
         - \frac{i}{2}\kappa_I{}^\bi{}_i\sigma^m\CD_m\bar\kappa_I{}^i{}_\bi
           \nonumber \\
    & &  + i\bar Q_{IA}{}^\bi{}_j D_I^{AB}{}^j{}_i Q_{IB}{}^i{}_\bi
         - i\bar Q_{IA}{}^\bj{}_i D^{AB}_{I+1}{}^\bi{}_\bj Q_{IB}{}^i{}_\bi
           \nonumber \\
    & &  - (\bar Q_{IA}{}^\bi{}_j\phi_I{}^j{}_i
         - \bar Q_{IA}{}^\bj{}_i\phi_{I+1}{}^\bi{}_\bj )
           (\bar\phi_I{}^i{}_kQ^A_I{}^k{}_\bi-\bar\phi_{I+1}{}^\bk{}_\bi Q^A_I{}^i{}_\bk)
           \nonumber \\
    & &  - (\bar Q_{IA}{}^\bi{}_j\bar\phi_I{}^j{}_i
         - \bar Q_{IA}{}^\bj{}_i\bar\phi_{I+1}{}^\bi{}_\bj )
           (\phi_I{}^i{}_kQ^A_I{}^k{}_\bi-\phi_{I+1}{}^\bk{}_\bi Q^A_I{}^i{}_\bk)
           \nonumber \\
    & &  + i\bar\psi_I{}^\bi{}_j\bar\lambda_I^A{}^j{}_i Q_{IA} {}^i{}_\bi
         - i\bar Q^A_I{}^\bi{}_j\lambda_{IA}{}^j{}_i \psi_I {}^i{}_\bi
         - i\bar\psi_I{}^\bj{}_i\bar\lambda^A_{I+1}{}^\bi{}_\bj Q_{IA} {}^i{}_\bi
         + i\bar Q^A_I{}^\bj{}_i\lambda_{I+1A}{}^\bi{}_\bj \psi_I {}^i{}_\bi
           \nonumber\\
    & &  + i\kappa_I{}^\bi{}_j\lambda^A_I{}^j{}_i Q_{IA} {}^i{}_\bi
         + i\bar Q^A_I{}^\bi{}_j\bar\lambda_{IA}{}^j{}_i \bar\kappa_I {}^i{}_\bi
         - i\kappa_I{}^\bj{}_i\lambda^A_{I+1}{}^\bi{}_\bj Q_{IA} {}^i{}_\bi
         - i\bar Q^A_I{}^\bj{}_i\bar\lambda_{I+1A}{}^\bi{}_\bj \bar\kappa_I {}^i{}_\bi
           \nonumber\\
    & &  + \frac{1}{\sqrt{2}}\kappa_I{}^\bi{}_j\phi_I{}^j{}_i  \psi_I {}^i{}_\bi
         + \frac{1}{\sqrt{2}}\bar\psi_I{}^\bi{}_j\bar\phi_I{}^j{}_i  \bar\kappa_I {}^i{}_\bi
%-\bar Q^A_I{}^\bi{}_i\CD^m\CD_m Q_{IA}{}^i{}_\bi
%-\bar Q_{IA}{}^\bj{}_i(\phi\bar\phi+\bar\phi\phi)_{I+1}{}_\bj{}^\bi Q^{A}_I{}^i{}_\bi
%\nonumber\\&&
%-\frac{i}{2}\bar\psi_I{}^\bj{}_i\bar\sigma^m\CD_m\psi_I{}^i{}_\bi
%-\frac{i}{2}\kappa_I{}^\bj{}_i\sigma^m\CD_m\bar\kappa_I{}^i{}_\bi
%\nonumber\\&&
         - \frac{1}{\sqrt{2}}\kappa_I{}^\bj{}_i\phi_{I+1}{}^\bi{}_\bj  \psi_I {}^i{}_\bi
         - \frac{1}{\sqrt{2}}\bar\psi_I{}^\bj{}_i\bar\phi_{I+1}{}^\bi{}_\bj \bar\kappa_I {}^i{}_\bi
           \Big]~.~~~
           \label{Sq comp}
    \end{eqnarray}

  By construction the action $S_V+S_q$ is invariant under the $\CN=2$ supersymmetry transformation law:
    \begin{eqnarray}
    \delta_\eta\phi_I
    &=&  - i\sqrt{2}\epsilon_{AB}\eta^A\lambda^B_I~,
           \\
%\delta_\eta\bar\phi&=&
%-i\sqrt{2}\epsilon_{ij}\bar\eta^i\bar\lambda^j~,\\
    \delta_\eta v_{Im}
    &=&    i\epsilon_{AB}(
           \eta^A\sigma_m\bar\lambda^B_I+\lambda^A_I\sigma_m\bar\eta^B
           )~,
           \\
           \delta_\eta\lambda^A_I{}_\alpha&=&
           \frac{1}{2}\sigma^m\bar\sigma^n\eta^Av_{Imn}
         + \sqrt{2}\sigma^m\bar\eta^A\CD_m\phi_I
         - i\eta^A[\phi_I,\bar\phi_I]
         + D_I{}^{A}{}_B\eta^B~,
           \label{delta lambda}
           \\
%\delta_\eta\bar\lambda^{i\dot\alpha}&=&
%\frac{1}{2}\bar\sigma^m\sigma^n\bar\eta^iv_{Imn}
%-\sqrt{2}\bar\sigma^m\eta^i\CD_m\bar\phi_I
%+i\bar\eta^i[\bar\phi_I,\phi_I]
%-D_I{}^{A}{}_B
%\bar\eta^B~,\\
    \delta_\eta D^{AB}_I
    &=&  - 2i\eta^A\sigma^m\CD_m\bar\lambda^B_I
         + 2i\CD_m\lambda^A_I\sigma^m\bar\eta^B
         + 2\sqrt{2}\bar\eta^A[\bar\lambda^B_I,\phi_I]
         + 2\sqrt{2}\eta^A[\lambda^B_I,\bar\phi_I]~,
           \\
    \delta_\eta Q^A_I{}^i{}_\bi
    &=&    \eta^A\psi_I{}^i{}_\bi+\bar\eta^A\bar\kappa_I{}^i{}_\bi~,\\
           \delta_\eta \psi_{I\alpha}{}^i{}_\bi&=&
           2i(\sigma^m\bar\eta^A)_\alpha\CD_mQ_{IA}{}^i{}_\bi
         - 2\sqrt{2}\eta^A_\alpha\bar\phi_I{}^i{}_j Q_{IA}{}^j{}_\bi
         + 2\sqrt{2}\eta^A_\alpha\bar\phi_{I+1}{}^{\bj}{}_\bi Q_{IA}{}^i{}_\bj~,
           \\
    \delta_\eta \bar\kappa_{I\dot\alpha}{}^i{}_\bi
    &=&    2i(\eta^i\sigma^m)_{\dot\alpha}\CD_mQ_{IA}{}^i{}_\bi
         + 2\sqrt{2}\bar\eta^A_{\dot\alpha}\phi_I{}^i{}_j Q_{IA}{}^j{}_\bi
         - 2\sqrt{2}\bar\eta^A_{\dot\alpha}\phi_{I+1}{}^\bj{}_\bi Q_{IA}{}^i{}_\bj~.
    \end{eqnarray}

%%%%%%%%%%%%%%%%%%%%%%%%%%%%%%%%%%%%%%%%%%%%%%%%%%%%%%%%%%%%
\subsection*{electric and magnetic FI terms}
  We introduce the electric FI term
    \bea
    S_e
     =     \int d u d \zeta^{(- 4)}
           \sum_{I=1}^n
           [ {\rm Tr}_{U(N_I)} \Xi^{++}_I V_I^{++}
         + h.c. ]
     =     \int d^4x \sum_{I=1}^n [\xi^{AB}_I D_{IAB}^0 + h.c. ]
           \label{Se}
    \eea
  where $\Xi^{++}_I = \xi^{AB}_I u_A^+u_B^+$ and $\xi^{AB}_I$ is the electric FI parameter of $U(N_I)$ gauge group.
  In the three vector notation, the electric FI parameter can be written as
  $\xi^A_{I B} = i \xi^\alpha_I (\tau_\alpha)^A_{~B}$
  where $\tau_\alpha$ ($\alpha = 1,2,3$) are the Pauli matrices.
  $S_e$ causes a constant shift of the auxiliary fields in the dual vector multiplets
  and thus the magnetic FI term $S_m$ is introduced to shift the auxiliary field 
  in the original vector multiplets by a constant \cite{APT, PP, Ivanov, FIS3}.
  We shall shift the auxiliary field as
    \bea
    \boldsymbol{D}_I^{aAB} 
    &=&   D_I^{aAB} + 4 i \xiD^{IAB} \delta^a_0~,
          ~~~~
    \bar{\boldsymbol{D}}_I^{aAB}
     =    D_I^{aAB} - 4 i \xiDb^{IAB} \delta^a_0~,
          \label{shift}
    \eea
  so that the supersymmetry transformation law (\ref{delta lambda}) changes to
    \begin{eqnarray}
    \delta \lambda_I^A{}^{a}
    &=&    (\boldsymbol{D}_I^a)^{A}_{~B} \eta^B + \ldots~,
           ~~~ 
    \delta \bar{\lambda}_I^{A}{}^a 
     =   - (\bar{\boldsymbol{D}}_I^a)^{A}_{~B} \eta^B + \ldots.
           \label{susytr shifted}
    \end{eqnarray}
  It is easy to see that the action with the shift (\ref{shift})
    \begin{eqnarray}
    S_V|_{D\to\boldsymbol{D}}
     =   - \frac{i}{4}\int d^4x  \sum_{I=1}^n [(D)^4 \CF_I(\hat W_I)
         - (\bar D)^4 \bar\CF_I(\bar{\hat W}_I)]~
           \label{SV shifted}
    \end{eqnarray}
  where $\hat W\equiv W |_{D\to\boldsymbol{D}}$
  is invariant under the $\CN=2$ supersymmetry transformations with (\ref{susytr shifted}).
  In addition obviously $S_e|_{D\to\boldsymbol{D}}$ is $\CN=2$ superinvariant.
  As we will see in the next section, the magnetic FI term
  introduced above causes partial spontaneous supersymmetry breaking.

  Next we examine $S_q$ in (\ref{Sq comp}).
  It is known \cite{PP, Ivanov} that
  there is a difficulty in introducing  the magnetic FI term
  in the presence of hypermultiplets in fundamental representation.
  However we find that the magnetic FI term can be introduced without breaking $\CN=2$ supersymmetry
  in the case with hypermultiplets in bi-fundamental representation.
  Let us show this explicitly.
  We examine the following terms contained in $S_q$
    \begin{eqnarray}
    & &    i\bar Q_{IA}{}^\bi{}_j D_I^{AB}{}^j{}_i Q_{IB}{}^i{}_\bi
         - i\bar Q_{IA}{}^\bj{}_i D^{AB}_{I+1}{}^\bi{}_\bj Q_{IB}{}^i{}_\bi
           \nonumber \\
    & &  + i\bar\psi_I{}^\bi{}_j\bar\lambda_I^A{}^j{}_i Q_{IA} {}^i{}_\bi
         - i\bar\psi_I{}^\bj{}_i\bar\lambda^A_{I+1}{}^\bi{}_\bj Q_{IA} {}^i{}_\bi
         - i\bar Q^A_I{}^\bi{}_j\lambda_{IA}{}^j{}_i \psi_I {}^i{}_\bi
         + i\bar Q^A_I{}^\bj{}_i\lambda_{I+1A}{}^\bi{}_\bj \psi_I {}^i{}_\bi
           \nonumber \\
    & &  + i\kappa_I{}^\bi{}_j\lambda^A_I{}^j{}_i Q_{IA} {}^i{}_\bi
         - i\kappa_I{}^\bj{}_i\lambda^A_{I+1}{}^\bi{}_\bj Q_{IA} {}^i{}_\bi
         + i\bar Q^A_I{}^\bi{}_j\bar\lambda_{IA}{}^j{}_i \bar\kappa_I {}^i{}_\bi
         - i\bar Q^A_I{}^\bj{}_i\bar\lambda_{I+1A}{}^\bi {}_\bj\bar\kappa_I {}^i{}_\bi~.
           \label{matter}
    \end{eqnarray}
  Under the shift (\ref{shift}), (\ref{matter}) acquires additional terms
    \begin{eqnarray}
    %\frac{i}{2} 
    -2\bar{Q}_{IA}{}^{\bi}{}_{i} 
    (\xiD^I-\xiDb^I)^{AB}(t_0^{I})^i{}_{j} 
    Q_{IB}{}^{j}{}_{\bi}
    %+\frac{i}{2} 
    +2\bar{Q}_{IA}{}^{\bi}{}_{i} 
    (\xiD^{I+1}-\xiDb^{I+1})^{AB}(t_0^{I+1})^\bj{}_{\bi} 
    Q_{IB}{}^{i}{}_{\bj}~.
    \label{additional}
    \end{eqnarray}
Now for the $\CN=2$ invariance of the action with the replacement (\ref{shift}),
  the following terms have to vanish
    \bea
    & &  - 2\delta\bar{Q}_{IA}{}^{\bi}{}_{i} 
           (\xiD^I-\xiDb^I)^{AB}(t_0^{I})^i{}_{j} 
           Q_{IB}{}^{j}{}_{\bi}
         - 2\bar{Q}_{IA}{}^{\bi}{}_{i} 
           (\xiD^I-\xiDb^I)^{AB}(t_0^{I})^i{}_{j} 
           \delta Q_{IB}{}^{j}{}_{\bi}
           \nonumber \\  
    & &  + 2\delta \bar{Q}_{IA}{}^{\bi}{}_{i} 
           (\xiD^{I+1}-\xiDb^{I+1})^{AB}(t_0^{I+1})^\bj{}_{\bi} 
           Q_{IB}{}^{i}{}_{\bj}
         + 2 \bar{Q}_{IA}{}^{\bi}{}_{i} 
           (\xiD^{I+1}-\xiDb^{I+1})^{AB}(t_0^{I+1})^\bj{} _{\bi}
           \delta Q_{IB}{}^{i}{}_{\bj}
           \nonumber \\  
    & &  - 4 \bar{\psi}_I{}^{\bi}{}_{i} 
           (\xiDb^{I}{}^A{}_{B} \bar{\eta}^B) (t^I_0)^i{}_{j} 
           Q_{IA}{}^{j}{}_{\bi}
         + 4 \bar{\psi}_I{}^{\bi}{}_{i} 
           (\xiDb^{I+1}{}^A{}_{B} \bar{\eta}^B) (t^{I+1}_0)^{\bj}{}_{\bi} 
           Q_{IA}{}^{i}{}_{\bj}
           \nonumber \\
    & &  + 4 \bar{Q}^A_I{}^{\bi}{}_{i} 
           (\xiD^I{}_{AB} \eta^B) (t^I_0)^i{}_{j} 
           \psi_I{}^{j}{}_{\bi}
         - 4 \bar{Q}^A_I{}^{\bi}{}_{i} 
           (\xiD^{I+1}{}_{AB} \eta^B) (t^{I+1}_0)^{\bj}{}_{\bi} 
           \psi_I{}^{i}{}_{\bj}
           \nonumber \\
    & &  - 4 \kappa_I{}^{\bi}{}_{i} 
           (\xiD^I{}^A{}_B \eta^B)(t^I_0)^i{}_{j} 
           Q_{IA}{}^{j}{}_{\bi}
         + 4 \kappa_I{}^{\bi}{}_{i} 
           (\xiD^{I+1}{}^{A}{}_{B} \eta^B) (t^{I+1}_0)^{\bj}{}_{\bi} 
           Q_{IA}{}^{i}{}_{\bj}
           \nonumber \\
    & &  - 4 \bar{Q}^A_{I}{}^{\bi}{}_{i} 
           (\xiDb^I{}_{AB} \bar{\eta}^B) (t^I_0)^i{}_{j} 
           \bar{\kappa}_I{}^{j}{}_{\bi}
         + 4 \bar{Q}^A_I{}^{\bi}{}_{i} 
           (\xiDb^{I+1}{}_{AB} \bar{\eta}^B) (t^{I+1}_0)^{\bj}{}_{\bi}
           \bar{\kappa}_I{}^{i}{}_{\bj}~.
           \label{mag FI susy}
    \eea
  We find that this is achieved if we choose the magnetic FI parameters such that 
    \bea
    \xiD^I
     =    \xiD^{I+1}
           \label{xi}
    \eea 
  where $(t^I_0)^i_{~j} = \delta^i_{~j}/\sqrt{2 N}$.
  A bi-fundamental hypermultiplet interacts with two different gauge sectors, 
  and thus we can introduce the magnetic FI terms 
  such that the effect from the shift of the auxiliary field of one gauge sector and that of the other sector 
  cancel out with each other.
  We can also see that the matter part does not contribute to the magnetic FI term
  as the additional terms (\ref{additional}) cancels out for (\ref{xi}).

  Summarizing the action of the $\CN=2$ quiver gauge model is given by
    \begin{eqnarray}
    S
    &=&    S_V+S_q+S_e+S_m
     =     \left[S_V+S_q+S_e \right] |_{D\to\boldsymbol{D}}~.
           \label{Sm}
    \end{eqnarray}
  Each part is given in (\ref{SV}), (\ref{Sq}) and (\ref{Se}), and $\xiD$ is subject to (\ref{xi}).
  We have seen that this is invariant under the $\CN=2$ supersymmetry transformation 
  with the replacement (\ref{shift}).

  We comment on the case with hypermultiplets in fundamental representation.
  As pointed out in \cite{PP, Ivanov},
  it is hard to introduce the magnetic FI term which causes
  the partial spontaneous supersymmetry breaking.
    \footnote{In \cite{FIS3}, the magnetic FI term is introduced even in the presence of hypermultiplets 
              in fundamental representation.
              However, as explained below,  the magnetic FI parameter is imaginary,
              and thus $\CN=2$ supersymmetry remains unbroken in the vacua.}
  This can be seen as follows.
  In this case, terms with $\xi_D^{I+1}=\bar \xi_D^{I+1}=0$ in (\ref{mag FI susy})
  have to be deleted for the $\CN=2$ superinvariance.
  This forces us to set the magnetic FI parameter $\xiD^{IAB}$ to be imaginary.
  However, the real part of the magnetic FI parameter causes partial spontaneous supersymmetry breaking \cite{APT},
  and thus for imaginary magnetic FI parameter $\CN=2$ supersymmetry remains unbroken in the vacua.
  In other words, 
  when we introduce the real part of the magnetic FI parameter
  in the presence of hypermultiplets in fundamental representation,
  the action is no longer invariant under the $\CN=2$ supersymmetry transformation.

  In this paper we mainly consider affine $A_{n-1}$ quiver gauge models.
  However, it is now obvious that we can introduce the magnetic FI term in any $\CN=2$ quiver gauge model 
  with any number of nodes (gauge sectors) and any number of arrows (bi-fundamental hypermultiplets).
  Let us comment on two generalizations of our model among them.
  The first one is the case when  the rank of each gauge group is different.
  Such an $\CN=2$ quiver gauge model is obtained 
  by considering the additional D5-branes wrapping on non-trivial $S^{2}$'s (though we have non-canonical kinetic terms)
  and is also interesting because, in contrast to the superconformal case above, we have running gauge couplings.
  Even in this case, the argument on the magnetic FI term is similar to that given above.
  The only difference is that the condition of the magnetic FI parameter (\ref{xi}) is changed as
    \bea
    \frac{\xiD^I}{\sqrt{N_I}}
     =    \frac{\xiD^{I+1}}{\sqrt{N_{I+1}}}~.
          \label{xi2}
    \eea 
  This change comes from the normalization of the generator: $(t^I_0)^i_{~j} = \delta^i_{~j}/\sqrt{2 N_I}$.
  So we have no difficulty in adding the magnetic FI term.
  The second one is the case when the gauge group is different from $A_{n-1}$.
  In fact, we can also construct $\CN=2$, $D_n$, $E_6$, $E_7$ and $E_8$ quiver gauge models.
  Even in these cases, all we have to do is to relate the FI parameters 
  in accordance with (\ref{xi}) or (\ref{xi2}).
  
%%%%%%%%%%%%%%%%%%%%%%%%%%%%%%%%%%%%%%%%%%%%%%%%%%%%%%%%%%%%%%%%%%%%%%%%
\section{The minimal model}

  We will show that in our model $\CN=2$ supersymmetry is partially broken to $\CN=1$ spontaneously.
  As an illustration we will focus on the affine $A_1$ quiver gauge model to which we refer as the minimal model
  in the following sections.
  
  The minimal model is composed of a pair of  hypermultiplets $q^+_I$
  and a pair of vector multiplets $V^{++}_I$ ($I=1,2$).
  $q^+_1$ and $q^+_2$, respectively, transform as bi-fundamental, $(\BN_1,\bar{\BN}_2)$ and $(\bar{\BN}_1,\BN_2)$, 
  and $V^{++}_I$ transform as adjoint under $U(N_I)$.
  The action of this model is given by (\ref{Sm}), with summing only over $I=1,2$.
  
  Let us write down the scalar potential in component.
  The scalar potential is
    \bea
    V
    &=&    \sum_{I=1,2} [ V_I^{(1)}+ V_I^{(2)} ]~,           
           \label{V} \\
    V_I^{(1)}
    &=&          \frac{1}{2} g_{Iab} \CP_I^a \CP_I^b
         + \frac{1}{4} g_{Iab} D_I^{a AB} | D_{I AB}^b |
         - 2 i \xiD^{I}{}^{AB} \xiD^I{}_{AB} \CF_{I|00} |
         + 2 i {\xiDb}^I{}^{AB} {\xiDb}^I{}_{AB} \bar{\CF}_{I|00} |
           \nonumber \\
    & &  - 4 i \xi_I^{AB} \xiD^I{}_{AB} 
         + 4 i \bar{\xi}_I^{AB} {\xiDb}^I{}_{AB}~,
           \label{V1}\\
    V_I^{(2)}   
    &= &   (\bar Q_{IA}{}^\bi{}_j \phi_I{}^j{}_i
         - \bar Q_{IA}{}^\bj{}_i \phi_{I+1}{}^\bi{}_\bj )
           (\bar\phi_I{}^i{}_kQ^A_I{}^k{}_\bi-\bar\phi_{I+1}{}^\bk{}_\bi Q^A_I{}^i{}_\bk)
           \nonumber \\
    & &  + (\bar Q_{IA}{}^\bi{}_j\bar\phi_I{}^j{}_i
         - \bar Q_{IA}{}^\bj{}_i\bar\phi_{I+1}{}^\bi{}_\bj )
           (\phi_I{}^i{}_kQ^A_I{}^k{}_\bi-\phi_{I+1}{}^\bk{}_\bi Q^A_I{}^i{}_\bk)~
           \label{V2}
    \eea
  where $\CP_I^a = - i f^a_{Ibc} \bar{\phi}^b_I \phi^c_I$, 
  $\CF_{I | ab\dots} |$ represents $\CF_{I | ab\dots}$ evaluated at $\theta^{\pm} = \bar{\theta}^{\pm} = 0$,
  and $g_{I | ab} = \Im \CF_{I | ab} |$ is the K\"ahler metric.
  $D^{a AB}_{I}$ is obtained by solving the equation of motion as
    \bea
    D_I^{a AB} |
     =   - 2 g_I^{ab} 
           [ (\xi_I + \bar{\xi}_I)^{AB} \delta^0_b 
         + \xi_{ID}^{AB} \CF_{I|0b} | + \bar{\xi}_{ID}^{AB} \bar{\CF}_{I|0b} | 
         + \CQ_{1I|b}^{AB} + \CQ_{2I|b}^{AB}]
         ~
    \eea
  where $\CQ_{1I}$ and $\CQ_{2I}$ are the contributions 
  from the hypermultiplets $q^+_1$ and $q^+_2$ respectively:
    \bea
    \CQ_{11|b}^{AB}
    &=&    \frac{i}{2} [\bar{Q}^A_1{}^\bi{}_{i} (t^1_b)^{i}_{~j} Q^{B j}_{1~~\bi}
         + (A \leftrightarrow  B)], 
           ~~~~    
    \CQ_{21|b}^{AB}
     =     -\frac{i}{2} [\bar{Q}^A_2{}^\bj{}_{i} (\bar{t}^1_b)^{\bi}{}_{\bj} 
     Q^{B i}_{2~~\bi}
         + (A \leftrightarrow  B)],
           \nonumber \\
    \CQ_{12|b}^{AB}
    &=&   - \frac{i}{2} [\bar{Q}^A_1{}^\bj{}_{i} (\bar{t}^2_b)^{\bi}{}_{\bj} 
    Q^B_1{}^i{}_{\bi}
         + (A \leftrightarrow  B)], 
           ~~~~    
    \CQ_{22|b}^{AB}
     =     \frac{i}{2} [\bar{Q}^{A \bi}_{2~i} (t^2_b)^{i}{}_{j} 
     Q^{B j}_{2~~\bi}
         + (A \leftrightarrow  B)]~.~~~
    \eea
  We can rewrite $V_I^{(1)}$ in  (\ref{V}) as 
    \bea
V_I^{(1)}&=&
    \frac{1}{2} g_{Iab} \CP_I^a \CP_I^b
    + \frac{1}{4} g_{Iab} \BD_I^{a AB} | \bar{\BD}_{I AB}^b |
    - 2 i (\xi_I^{AB} - \bar{\xi}_I^{AB})(\xiD^I{}_{AB} + {\xiDb}^I{}_{AB})
    \label{V1'}
    \eea
  where 
    \bea
    \boldsymbol{D}_I^{a AB} |
    &=&    D_I^{a AB} | + 4 i \xiD^I{}^{AB} \delta^a_0
           \nonumber \\
    &=&  - 2 g_I^{ab} 
           \left[~
           (\xi_I + \bar{\xi}_I)^{AB} \delta^0_b 
         + (\xiD^I + \xiDb^I)^{AB} \bar{\CF}_{I|0b} | 
         + \CQ_{1I|b}^{AB} + \CQ_{2I|b}^{AB}
           ~\right]
         ~.
    \eea
  
%%%%%%%%%%%%%%%%%%%%%%%%%%%%%%%%%%%%%%%%%%%%%%%%%%%%%
\section{Vacua of the minimal model}
  In this section, we will find the $\CN=1$ supersymmetric vacua in the Coulomb branch $\langle Q_I \rangle=0$
  by analyzing the condition stabilizing the scalar potential derived in the previous section.
  
  The constraint, $\langle g_{Iab} \CP_I^a \CP_I^b \rangle = 0$, can be satisfied 
  by vanishing non-diagonal components of the vacuum expectation value of $\phi_I$, that is, 
  $\langle \phi_I^r \rangle = 0$ where $t^I_{r}$ represent non-Cartan generators of the gauge group $U(N_I)$.
  Then, we consider the condition to stabilize the scalar potential.
  While the derivative of the scalar potential $V$ with respect to the hypermultiplet scalar $Q$ is trivially zero
  in the Coulomb branch $\langle Q_I \rangle=0$,
  the non-trivial vacuum condition is derived 
  from the derivative with respect to $\phi_I$ in the vector multiplet
    \bea
    0 
     =     \langle \frac{\partial}{\partial \phi^a_I} V \rangle
     =     \frac{i}{4} \sum_\alpha 
           \langle \CF_{I|abc} | \BD_I^{b \alpha} \BD_I^{c \alpha} \rangle,
    \eea
  where $ \BD_{I~B}^{a A} = i \textrm{\boldmath $D$}_I^{a \alpha} (\tau_\alpha)^A_{~B} $.
  Note that the index $I$ is not summed over here.
  
  Let us examine the case with the single trace prepotential of degree $n_I + 2$
    \bea
    \CF_I
     =     \sum_{k=1}^{n_I + 1} \frac{g_k}{(k+1)!} \Tr W^{k + 1}
    \eea
  for concreteness.
  Let  $E^I_{\ui \uj}$, $\ui = \underline{1}, \ldots, \underline{N_I}$,
  be the fundamental matrix of gauge group $U(N_I)$
  which has $1$ at the $(\ui, \uj)$ component and $0$ otherwise.
  Cartan generators can be written as $t^I_{\ui} = E^I_{\ui \ui}$.
  We have $\langle \partial V/ \partial \phi_I^r \rangle =0$ 
  because $\langle \CF_{I | r \ui \ui} \rangle = \langle \BD_I^r \rangle = 0$.
  Noting that the points specified by $\langle \CF_{I | \ui \ui \ui} \rangle =0$ correspond to the unstable vacua, 
  we derive the vacuum conditions as follows,
    \bea
    \sum_\alpha \langle \BD_I^{\ui \alpha} \BD_I^{\ui \alpha} \rangle
     =     0, 
           ~~~~~~~
           {\rm for~all}~~\ui~{\rm and}~I.
    \eea
  As in \cite{FIS3}, we can choose the FI parameters by using $SU(2)$ rotation as
    \begin{eqnarray}
    (\xi_{I} + \bar{\xi}_{I})^\alpha
     =     (0, e_I, \xi_I)~,~~~
    (\xiD^{I} + {\xiDb}^{I})^\alpha
     =     (0, m_I, 0)~.
           \label{FIpara}
    \end{eqnarray}
  Furthermore, we can set $m_I / \xi_I < 0$, without loss of generality.
  In these choices of the FI parameters, we obtain the following vacuum condition, 
    \bea
    \langle \CF_{I | \ui \ui} \rangle
     =   - 2 \left( \frac{e_I}{m_I} + i \frac{\xi_I}{m_I} \right).
    \eea
  Note that the minus sign in front of $i \xi_I / m_I$ has been excluded 
  by the positivity criterion of the K\"ahler metric: 
  $\langle g_{I| \ui \ui} \rangle = \Im \langle \CF_{I | \ui \ui} \rangle > 0$.
  In the original bases, this means
    \bea
    \langle \CF_{I | 00} \rangle
     =   - \left( \frac{e_I}{m_I} + i \frac{\xi_I}{m_I} \right).
    \eea
  The vacuum expectation values of the diagonal components of $\phi_I$ are determined 
  from the above equations of degree $n_I$. 
  Thus, the gauge symmetry $U(N_I)$ is broken to ${\displaystyle \prod_{i=1}^{n_I}} U(N_{I|i})$ 
  with $N_I = {\displaystyle \sum_{i=1}^{n_I}} N_{I|i}$.
  
  We can easily evaluate the vacuum energy
      \bea
    \langle V \rangle
     =     \sum_{I=1,2}
           \left(
         - 4 m_I \xi_I
         - 4 i \sum_\alpha (\xi_I - \bar{\xi}_I)^\alpha (\xiD^I + {\xiDb}^I)^\alpha
           \right)
    \eea
  which comes from the last two terms in (\ref{V1'}).
  As pointed out in \cite{APT}, using the freedom to choose the imaginary part of $\xi_I^\alpha$, 
  we can obtain the vanishing vacuum energy; 
  if we set $(\xi_I - \bar{\xi}_I)^2 = i \xi_I$, then the vacuum energy is zero.
  The vanishing vacuum energy may indicate that $\CN=1$ supersymmetry remains in the vacuum.

  In the subsequent subsections, we will show that 
  the mass spectrum on the vacuum can be written in terms of $\CN=1$ multiplets, 
  and that a linear combination of fermions becomes the Nambu-Goldstone fermion
  associated with the partial supersymmetry breaking.

%%%%%%%%%%%%%%%%%%%%%%%%%%
\subsection{Mass spectrum}
  The masses of the component fields contained in the vector multiplets are similar to 
  those in the pure $U(N)$ Yang-Mills case \cite{FIS1, FIS2}.
  The $\CN=2$, $U(N_I)$ vector multiplet decomposes into three types of $\CN=1$ multiplets:
  $\CN=1$, ${\displaystyle \prod_{i=1}^{n_I}} U(N_{I|i})$ massless vector multiplet, 
  $\CN=1$ massive chiral multiplet in adjoint representation with mass 
  $M_I=m_I\langle g_I^{\alpha\alpha}\CF_{I|0\alpha\alpha}\rangle$
  where $t_\alpha^I$ represent unbroken generators,
  and $\CN=1$ massive vector multiplets which correspond to the broken generators when the gauge symmetry is broken.
  
  Let us turn to the matter part.
  The mass of the scalar components $Q_{IA}{}^i{}_\bi$ is easily obtained by evaluating the second derivative of $V$.
  First we examine $V_I^{(1)}$ in (\ref{V1'}).
  We observe its vacuum expectation value vanishes as follows.
  Since the first and the last terms in (\ref{V1'}) do not contain the scalar $Q_I$,
  only the second term can contribute to the mass.
  However, it vanishes as
    \bea
    \left<
    \partial_{\bar{Q}_{J~i}^{A\bi}} \partial_{Q_{K~\bj}^{Bj}}
    \sum_I g_{I | ab}
    \BD_I^{a CD} | \bar{\BD}_{I CD}^b | 
    \right>
    &=&  - 4 \sum_I \left<
           ( \Re \BD_{ICD}^{a} )
           \partial_{\bar{Q}_{J~i}^{A\bi}} \partial_{Q_{K~\bj}^{Bj}}
           (\CQ_{1I | a}^{CD} + \CQ_{2I | a}^{CD})
           \right>
           \nonumber \\
    &=&  - 4 i \delta^1_J \delta^1_K
           \left<
           (t_a^1)^i_{~j} \delta_{~\bi}^{\bj} \Re \BD_{1AB}^{a}
         - (\bar{t}_a^2)_{~\bi}^{\bj} \delta^i_{~j} \Re \BD_{2AB}^{a}
           \right>
           \nonumber \\
    & & - 4 i \delta^2_J \delta^2_K
           \left<
          -(\bar{t}_a^1)^{i}_{~j} \delta_{~\bi}^{\bj} \Re \BD_{1AB}^{a}
         + (t_a^2)_{~\bi}^{\bj} \delta^{i}_{~j} \Re \BD_{2AB}^{a}
           \right>
           \nonumber \\
    &=&    0.
    \eea
  In the last equality, we have used 
  $\langle \Re \BD_{IAB}^{a} \rangle = - 2 i (\tau_2)_{AB} m_I \delta^a_0$
  and the relation (\ref{xi}).
  Next, we examine $V_I^{(2)}$ in (\ref{V2}).
  It is easy to see that the scalar components $Q_{1A~\bj}^{~i}$ and $Q_{2A~\bj}^{~i}$
  have the same mass $m^2_{i \bj} = 2 |a_{1 i} -a_{2 \bj}|^2$ where
    \begin{eqnarray}
    \langle \phi_I \rangle
    &=&    \mathrm{diag} (a_{I1}, \ldots ,a_{I N_I})~.
    \label{vev of phi}
    \end{eqnarray}
  
  The masses of the fermions $\kappa^I$ and $\psi^I$ can be seen from the following terms of $S_q$ (\ref{Sq comp})
    \begin{eqnarray}
    \frac{1}{\sqrt{2}} \kappa_1{}^{\bi}{}_{i} 
    (\phi_1{}^i{}_{j} \delta_{~\bi}^{\bj} - {\phi}_2{}_{~\bi}^{\bj} \delta^i_{~j})
     \psi_1{}^j{}_{\bj}
    + \frac{1}{\sqrt{2}} \kappa_2{}^{\bi}{}_{i} 
    (\phi_2{}^i{}_{j} \delta_{~\bi}^{\bj} - {\phi}_1{}_{~\bi}^{\bj} \delta^i_{~j})
     \psi_2{}^j{}_{\bj}
    + h.c.
    ~.
    \label{mass term: kappa psi}
    \end{eqnarray}
  Let us examine the first term from which masses of $\kappa_1$ and $\psi_1$ are determined.
  As $\langle \phi_I\rangle$ is diagonal (\ref{vev of phi}),
  we can rewrite it as
    \bea
    & &    \frac{1}{2 \sqrt{2}} 
           \left(
           \begin{array}{cc}
           \kappa^{~\bi}_{1~i} & \psi^{~i}_{1~\bi}
           \end{array}
           \right)
           \left(
           \begin{array}{cc}
           0 & a_{1i} - a_{2 \bj}  \\
           a_{1i} - a_{2 \bj}   & 0
           \end{array}
           \right)
           \left(
           \begin{array}{c}
           \kappa^{~\bi}_{1~i} \\
           \psi^{~i}_{1~\bi}
           \end{array}
           \right)
           \nonumber \\
    & &
     =     \frac{1}{2 \sqrt{2}} 
           \left(
           \begin{array}{cc}
           \psi_{+1}{}^{\bi}{}_{i} & \psi_{-1}{}^i{}_{\bi}
           \end{array}
           \right)
           \left(
           \begin{array}{cc}
           a_{1i} - a_{2 \bj}  & 0 \\
           0  & a_{1i} - a_{2 \bj} 
           \end{array}
           \right)
           \left(
           \begin{array}{c}
           \psi_{+1}{}^{\bi}{}_{i}  \\
           \psi_{-1}{}^i{}_{\bi}
           \end{array}
           \right)
           ~~~~~~
    \eea
  where we have defined $\psi_{\pm} \equiv (\kappa \pm \psi)/\sqrt{2}$.
  By taking the normalization of the kinetic terms into account,
  one sees that
  the masses of $\psi_{+ 1~\bj}^{~i}$ and $\psi_{- 1~\bj}^{~i}$ can be evaluated as $\sqrt{2 |a_{1 i} -a_{2 \bj}|^2}$.
  In the same way, by examining the second term in (\ref{mass term: kappa psi}),
  the mass of $\psi_{+ 2~\bj}^{~i}$ and $\psi_{- 2~\bj}^{~i}$ is found to be the same as that of $\psi_{\pm 1}$.
  
  Thus in the vacuum, $\CN=2$ hypermultiplet $q_1^+$ in $(\BN_1, \bar{\BN}_2)$ representation
  is decomposed into various massive multiplets according to the branching rule
  and the massive multiplets with mass $\sqrt{2 |a_{1 i} -a_{2 \bj}|^2}$
  transform as $( \BN_{1|i}, \bar{\BN}_{2|\bj})$. 
  The same is true for another hypermultiplet $q_2^+$ in $(\bar{\BN}_1, \BN_2)$.
  
%%%%%%%%%%%%%%%%%%%%%%%%%%
\subsection{Nambu-Goldstone fermion}
  In the case of the $\CN=2$, $U(N)$ gauge model with/without hypermultiplets in adjoint representation
  \cite{FIS1, FIS2, FIS3}, 
  a linear combination of the overall $U(1)$ fermions in the $\CN=2$ vector multiplet becomes the Nambu-Goldstone fermion.
  One might think that as there are two gauge sectors, $U(N)^2$, 
  two Nambu-Goldstone fermions would emerge.
  However, this is not correct.
  We show that only one combination of these fermions becomes the Nambu-Goldstone fermion.
  
  The vacuum expectation values of the supersymmetry transformations of component fields vanish 
  except for $\langle \delta \lambda_I^A\rangle$.
  In the choice of the FI parameters (\ref{FIpara}), we obtain
    \begin{eqnarray}
    \langle\boldsymbol{D}_I^a{}^A{}_B\rangle
     =     i m_I \delta^a_0
           \left(
           \begin{array}{cc}
           1   & -1   \\
           1   & -1   \\
           \end{array}
           \right)~
           \end{eqnarray}
  with $m_1 = m_2$ which follows from the condition (\ref{xi}).
  Thus, letting $\lambda^{a\pm}_I\equiv\frac{1}{\sqrt{2}}(\lambda^{a 1}_I\pm\lambda^{a2}_I)$,
    \begin{eqnarray}
    \langle \delta \lambda_I^{a+} \rangle
    &=&    i \sqrt{2} m_I\delta^a_0(\eta^1-\eta^2)~,
           ~~~
    \langle\delta \lambda_I^{a-}\rangle
     =     0~.
    \end{eqnarray}
  Furthermore, combining $\lambda_1^{0+}$ and $\lambda_2^{0+}$, we find
    \begin{eqnarray}
    \langle\delta (\lambda_1^{0+}-\lambda_2^{0+})\rangle
    &=&    i\sqrt{2} (m_1-m_2)(\eta^1-\eta^2)
     =     0~,
           \nonumber \\
    \langle\delta (\lambda_1^{0+}+\lambda_2^{0+})\rangle
    &=&    i \sqrt{2} (m_1+m_2)(\eta^1-\eta^2)
     \neq 0~
    \end{eqnarray}
  where in the last equality we have used $m_1 = m_2$.
  So, the fermion $\lambda_1^{0+}+\lambda_2^{0+}$ can be the Nambu-Goldstone fermion.
  In order to conclude that this is the Nambu-Goldstone fermion, 
  we have to show this fermion is exactly massless.

  The mass term of $\lambda^0$ can be read off from (\ref{SV:component}) as
    \begin{eqnarray}
    \frac{i}{4}\langle\CF_{I000}\boldsymbol{D}^{0AB}_I\rangle\lambda_I^{0}{}_A\lambda_I^{0}{}_B~,
    \end{eqnarray}
  with the replacement (\ref{shift}).
  It is easy to see that this is proportional to $im_I\lambda^{0-}_I\lambda^{0-}_I$,
  and thus we conclude that $\lambda_1^{0+}+\lambda_2^{0+}$ is massless as $\lambda^{0+}_I$ are massless.
  As a result, we can identify $\lambda_1^{0+}+\lambda_2^{0+}$ 
  as the Nambu-Goldstone fermion associated with the partial spontaneous supersymmetry breaking.
  
  Let us comment on the affine $A_{n-1}$ quiver gauge model with $n>2$.
  In this case $\delta \lambda_I^{0+}$ are independent of $I$ because of the condition (\ref{xi}).
  Therefore, only the vacuum expectation value of the supersymmetry transformation of
  the combination, $\tilde{\lambda} = \sum_I \lambda_I^{0+}$, is not zero.
  As $\lambda_I^{0+}$ are massless and so $\tilde\lambda$ are,
  we may identify $\tilde\lambda$ with the Nambu-Goldstone fermion 
  associated with partial supersymmetry breaking.

%%%%%%%%%%%%%%%%%%%%%%%%%%%%%%%%%%%%%%%%%%%%%%%%%%%%%%%%%%%%%%%%%%%%%%%%%%%%%%%%%%%%%%%%
\section{A dynamical realization of Klebanov-Witten model}
  In \cite{Klebanov:1998hh}, Klebanov and Witten considered 
  the $\CN=2$ affine $A_1$ quiver gauge theory
  realized on the world volume on D3-branes at orbifold singularity
  (dual to type IIB superstring in $AdS_5 \times S^5/\bZ_2$ \cite{Kachru:1998ys}),
  and discussed a flow to $\CN=1$ fixed point  by adding the mass operator \cite{Hanany:1998it}
  of the  chiral multiplet $\Phi_I$ in adjoint representation, which breaks $\CN=2$ supersymmetry to $\CN=1$.
  It was shown that the superpotential of the effective theory at the fixed point
  can be regarded as that of the world volume theory at the conifold singularity
  (dual to type IIB superstring in $AdS_5\times T^{1,1}$).

  As our minimal model is $\CN=2$ affine $A_1$ quiver gauge theory
  with non-canonical kinetic terms and electric and magnetic FI terms,
  it is expected that our model might describe the $\CN=1$ quiver gauge theory 
  dual to type IIB superstring in $AdS_5\times T^{1,1}$ in some points of vacua.
  If so, the minimal model may provide a dynamical realization of the statement of \cite{Klebanov:1998hh}
  because the $\CN=1$ chiral multiplet $\Phi_I$ in adjoint representation becomes massive dynamically.
  We show that this is the case.
  
  First, we examine the matter content realized on the vacuum.
  Let us consider the following point of vacua
    \begin{eqnarray}
    \langle \phi_1 \rangle
     =     \langle \phi_2 \rangle
     =     \diag (a,\ldots, a) ~.
           \label{VEV}
    \end{eqnarray}
  The $U(N)^2$  gauge symmetry is unbroken in this vacuum,
  while $\CN=1$ chiral multiplet in the adjoint representation becomes massive with  mass 
  $M_I = m_I \langle g_I^{a a} \CF_{I | 0 a a} \rangle$ \cite{FIS2}.
  As was seen in section 4.1,
  the $\CN=1$ chiral multiplets $(A_I,\bar B_I)$  in bi-fundamental representation
  acquire masses $m^2_{i \bj} = 2 |a_{1 i} -a_{2 \bj}|^2$,
  but in the vacuum (\ref{VEV}) these become massless.
  Thus, the massless multiplets in the vacuum are
  $\CN=1$, $U(N)^2$ vector multiplet, $\CN=1$ chiral multiplets $A_I$ and $B_I$ ($I=1,2$) in $(\BN,\bar\BN)$
  and $(\bar \BN, \BN)$ representations, respectively.
  This is exactly the matter content of the $\CN=1$ quiver gauge theory
  dual to IIB superstring in $AdS_5\times T^{1,1}$.
  
  Next, we examine the superpotential realized in the vacuum.
  Expanding each field around its vacuum expectation value, we obtain the fluctuation action in the $\CN=1$ vacuum.
  The matter part of the action can be written in terms of $\CN=1$ multiplets.
  The superpotential term which contains the  chiral multiplets $A_I$ and $B_I$ in bi-fundamental representation is
    \bea
    W_{matter}
     =     \Tr [(A_1 B_1 - A_2 B_2) \Phi_1 - (B_1 A_1 - B_2 A_2) \Phi_2]
    \eea
  up to an overall numerical coefficient.
  Here we have not written the gauge index explicitly.
  Of course, this is the ordinary superpotential of $\CN=2$ supersymmetric gauge theory with  hypermultiplets 
  in $\CN=1$ superspace formalism  \cite{Klebanov:1998hh, Hanany:1998it},
  because we are considering the Coulomb branch.
  On the other hand, the mass term of the chiral multiplets $\Phi_I$ in the superpotential is found to be
    \bea
    W_{\Phi}
     =    M_1 \Tr \Phi_1^2 + M_2 \Tr \Phi_2^2 ~.
    \eea
  The effective theory is described by the massless fields, 
  and the massive chiral multiplets $\Phi_I$ in adjoint representation should be integrated out.  
  To achieve this, we note that the mass matrix $M_I$ is determined 
  by the vacuum expectation value $\langle\phi\rangle$ and the prepotential $\CF_I$.
  Though we are working in special points of vacua,
  $M_I$ still depends on the choice of the prepotential function.
  Suppose that the prepotentials are set to satisfy $M_1 = - M_2$.
  By integrating out $\Phi_I$ using the equation of motion,
  we obtain the following superpotential up to an overall numerical coefficient
    \bea
    \Tr (A_1 B_1 A_2 B_2 - B_1 A_1 B_2 A_2)
    \label{T11}
    \eea
  which is the well-known superpotential of the world volume theory on the D3-branes at the conifold singularity.
  
  It is interesting to take the higher order terms in $\Phi_I$
  into account, and examine the deformations of the superpotential (\ref{T11}). 

%%%%%%%%%%%%%%%%%%%%%%%%%%%%%%%%%%%%%%%%%%%%%%%%%%%%%%%%%%%%%%%%%%%%%%%%%%%%%%%%%%%%%%%%
\section*{Acknowledgements}
  We thank Yosuke Imamura, Kazutoshi Ohta and Yukinori Yasui for useful discussion.
  This work is supported in part by the Grant-in-Aid for Scientific Research (18540285,
  19540324,
  19540304,
  19540098) 
  from the Ministry of Education, Science and Culture, Japan.
  Support from the 21 century COE program ``Constitution of wide-angle mathematical basis focused on knots'' 
  is gratefully appreciated.
  We thank the Yukawa Institute for Theoretical Physics at Kyoto University, 
  where the preliminary version of this work was presented, 
  during the YITP-W-07-05 on ``String Theory and Quantum Field Theory'' (6th-10th August 2007).
  We wish to acknowledge the participants for stimulating discussions.

%%%%%%%%%%%%%%%%%%%%%%%%%%%%%%%%%%%%%%%%%%%%%%%%%%%%%%%%%%%%%%%%%%%%%%%%%%%%%%%%%%%%%%%%
\appendix

\section*{Appendix}

\section{Superfields in Harmonic Superspace}
  Harmonic superspace \cite{HS, HSbook}
   is an extension of the $\CN=2$ superspace \cite{Grimm:1977xp} by including harmonic variables $u_A^{\pm}$ 
  ($A = 1, 2$)
    \begin{eqnarray}
    (u_A^+,u_A^-) \in SU(2)~,~~~
    u^{+A}u^-_A=1~,~~~
    \overline{u^{+A}}=u^-_A~
        \end{eqnarray}
  where the bar ``${}^-$" means the complex conjugation
    \bea
    \overline{Q^A}
    &=&    \bar{Q}_A,
           \nonumber \\
    \overline{Q_A}
    &=&    \overline{\varepsilon_{AB} Q^B}
     =     \epsilon_{AB} \bar{Q}_B
     =     \epsilon_{AB} \epsilon_{BC} \bar{Q}^C
     =   - \bar{Q}^A,
           \nonumber \\
    \overline{\xi^{AB}}
    &=&    \bar{\xi}_{AB},
           \nonumber \\
    \overline{\xi_{AB}}
    &=&    \overline{\epsilon_{AC} \epsilon_{BD} \xi^{CD}}
     =     \epsilon_{AC}  \epsilon_{BD} \bar{\xi}_{CD}
     =     \epsilon_{AC}  \epsilon_{BD} \epsilon_{CE}  \epsilon_{DF} \bar{\xi}^{EF}
     =     \bar{\xi}^{AB}.
    \eea
 
  We introduce an $\CN=2$ vector multiplet $V^{++}=V^{++a}t_a$
  transforming as adjoint representation under the gauge group:
  hermitian matrices $(t_a)^i{}_j$
% ($a=0,1,...,N_c^2-1$)
generate 
%$u(N_c)$, 
$[t_a,t_b]=if^a_{bc}t_a$.
$V^{++}$ is the analytic superfield satisfying
$D^+V^{++}=\bar D^+V^{++}=0$.
%and $t_0$ represents the overall $u(1)$ generator.
In the analytic basis
\begin{eqnarray}
&&
(y^m,\theta^\pm,\bar\theta^\pm,u_A^\pm)
%\nonumber\\&&~~
=
(x^m-2i\theta^{A}\sigma^m\bar\theta^{B}u_{(A}^+u_{B)}^-,\theta^Au_A^\pm,\bar\theta^Au_A^\pm,u_A^\pm)~,
\label{AB}
\end{eqnarray}
% $(x_A,\theta^\pm,\bar\theta^\pm,u^\pm_i)$ in
%(\ref{AB}),
$D^\pm$ and $\bar D^\pm$ are given as
\begin{eqnarray}
&&
D^+_\alpha=\frac{\partial}{\partial\theta^{-\alpha}}~,~~
D^-_\alpha=-\frac{\partial}{\partial\theta^{+\alpha}}
+2i(\sigma^m\bar\theta^{-})_\alpha\frac{\partial}{\partial y^m}~,
\nonumber\\&&
\bar D^+_{\dot \alpha}=\frac{\partial}{\partial\bar\theta^{-\dot\alpha}}~,~~
\bar D^-_{\dot \alpha}=
-\frac{\partial}{\partial\bar\theta^{+\dot\alpha}}
-2i(\theta^{-}\sigma^m)_{\dot \alpha}\frac{\partial}{\partial y^m}~,~~~
\end{eqnarray}
and thus $V^{++}$
is a superfield
in analytic subspace
$(y,\theta^+,\bar\theta^+,u^\pm_i)$~.
In the Wess-Zumino gauge
$V^{++}$ is given as
\begin{eqnarray}
V^{++}&=&-2i\theta^+\sigma^m\bar\theta^+v_m(y)
-i\sqrt{2}(\theta^+)^2\bar\phi(y)
%\nonumber\\&&~~
+i\sqrt{2}(\bar\theta^+)^2\phi(y)
+4(\bar\theta^+)^2\theta^+\lambda^A(y)u_A^-
\nonumber\\&&
-4(\theta^+)^2\bar\theta^+\bar\lambda^A(y)u_A^-
+3(\theta^+)^2(\bar\theta^+)^2D^{AB}(y)u_A^-u_B^-~,
\label{V++}
\end{eqnarray}
where $v_m$, $\phi$, $\lambda^A$ and $D^{AB}$
are vector, complex scalar, $SU(2)$ doublet Weyl spinors and 
auxiliary field, respectively.
%$D^{ij}$ is symmetric with respect to $(i,j)$ so that
%$D^i{}_{j}=\varepsilon_{jk}D^{ik}$
%is an $SU(2)$ matrix, $D^i{}_{j}=iD^A(\tau_A)^i{}_{j}$.
%The reality $V^{++}=\widetilde{V^{++}}$,
%where 
%the tilde
%``$\widetilde{~~~}$'' 
%means the analyticity preserving 
%conjugation \cite{HS} (see appendix A), implies that
%$D^{ij}=\bar D^{ij}$
%because
%$\overline{D^{ij}u_i^-u_j^-}=\bar D_{ij}u^{-i}u^{-j}=
%\bar D^{ij}u^{-}_iu^{-}_j$,
%and that the $D^A$ is a real three vector
%$
%\overline{D^A}=D^A~
%$.
The curvature $\bar{W}$ of $V^{++}$ defined by
\begin{eqnarray}
\bar{W}&=&
-\frac{1}{4}(D^+)^2
\sum_{n=1}^{\infty}
\intd v_1\cdots dv_n(-i)^{n+1}
%\nonumber\\&&\times
\frac{V^{++}(v_1)\cdots V^{++}(v_n)}
{(u^+v_1^+)(v_1^+v_2^+)\cdots(v_n^+u^+)}~
\end{eqnarray}
is evaluated to give \cite{FIS3}
\begin{eqnarray}
\bar W&=&
\bar\theta^A\bar\sigma^{mn}\bar\theta_A\, v_{mn}
-i\sqrt{2}\bar \phi
+i\sqrt{2}(\bar\theta)^4\eta^{mn}\CD_{m}\CD_n\phi
+\frac{4}{3}i(\bar\theta^A\bar\theta^B)\,\CD_m\lambda_A\sigma^m\bar\theta_B
-2\bar\theta^A\bar\lambda_A
\nonumber\\&&
+\bar\theta^A\bar\theta^BD_{AB}
-\frac{2}{3}\sqrt{2}(\bar\theta^A\bar\theta^B)[\phi, \bar\theta_A\bar\lambda_B]
+i(\bar\theta)^4\,\varepsilon_{AB}[\lambda^A, \lambda^B]
+i\sqrt{2}(\bar\theta)^4[\phi,[\phi,\bar \phi]]
\nonumber\\&&
{-2i\bar\theta^+\bar\theta^- [\phi,\bar\phi]}
+\cdots
\label{W}
\end{eqnarray}
where
$
v_{mn}=\partial_mv_n-\partial_nv_m+i[v_m,v_n]$ and $
\CD_m \circ=\partial_m\circ+i[v_m,\circ]
%\\
%\CD_m\lambda&=&\partial_m\lambda+i[v_m,\lambda]~.
$.
The ellipsis represents
terms which do not contribute to the action (\ref{SV}).

Next we introduce an
$\CN=2$ hypermultiplet $q^+{}^i{}_\bj$ transforming as bi-fundamental representation
under the gauge group.
The 
$q^+$ hypermultiplet
 is an analytic superfield
satisfying
$D^+q^{+}=\bar D^+q^{+}=0$,
and can be expanded as
\begin{eqnarray}
q^{+ }&=&
F^{+ }(y,u)
+\theta^{+}\psi (y,u)
+\bar\theta^+\bar\kappa (y,u)
%\nonumber\\&&
+(\theta^+)^2M^{- }(y,u)
+(\bar\theta^+)^2N^{- }(y,u)
\nonumber\\&&
+i\theta^+\sigma^m\bar\theta^+A_m^{- }(y,u)
+(\theta^+)^2\bar\theta^+\bar\gamma^{(-2) }(y,u)
%\nonumber\\&&
+(\bar\theta^+)^2\theta^+\chi^{(-2) }(y,u)
\nonumber\\&&
+(\theta^+)^2(\bar\theta^+)^2P^{(-3) }(y,u)~~~~~
\end{eqnarray}
in the analytic basis.
$q^+$ contains infinitely many auxiliary fields
which are eliminated by solving the equations of motion derived from (\ref{Sq}):
$\SD^{++}q^+=0$
where $\SD^{++}$ is given in (\ref{Dq})
and
\begin{eqnarray}
D^{++}&=&
\partial^{++}
-2i\theta^+\sigma^m\bar\theta^+\frac{\partial}{\partial y^m}
+\theta^{+\alpha}\frac{\partial}{\partial \theta^{-\alpha}}
+\bar\theta^{+\dot\alpha}\frac{\partial}{\partial \bar\theta^{-\dot\alpha}}
\,,~~~
\partial^{++}=u^{+A}\frac{\partial}{\partial u^{-A}}
~.
\end{eqnarray}
We find that the auxiliary fields are eliminated by
\begin{eqnarray}
F^+_I{}^i{}_\bi&=&Q^A_I(y){}^i{}_\bi u_A^+~,\\
\psi{}^i{}_\bi&=&\psi(y){}^i{}_\bi~,~~~
\bar\kappa{}^i{}_\bi=\bar\kappa(y){}^i{}_\bi~,\\
M^-_I{}^i{}_\bi&=&-\sqrt{2}(\bar\phi_I{}^i{}_jQ^A_I{}^j{}_\bi
-\bar\phi_{I+1}{}^\bj{}_\bi Q^A_I{}^i{}_\bj)u_A^-~,\\ 
N^-_I{}^i{}_\bi&=&\sqrt{2}(\phi_I{}^i{}_jQ^A_I{}^j{}_\bi
-\phi_{I+1}{}^\bj{}_\bi Q^A_I{}^i{}_\bj)u_A^-~,\\
A_{Im}^-{}^i{}_\bi&=&
2\CD_mQ^A_I{}^i{}_\bi u_A^-~,\\
\bar\gamma^{(-2)}_I{}^i{}_\bi&=&
2i(\bar\lambda^A_I{}^i{}_jQ^B_I{}^j{}_\bi
-\bar\lambda^A_{I+1}{}^\bj{}_\bi Q^B_I{}^i{}_\bj
)u_A^-u_B^-~,\\
\chi^{(-2)}_I{}^i{}_\bi&=&
-2i(\lambda^A_I{}^i{}_jQ^B_I{}^j{}_\bi
-\lambda^A_{I+1}{}^\bj{}_\bi Q^B_I{}^i{}_\bj
)u_A^-u_B^-~,\\
P^{(-3)}_I{}^i{}_\bi&=&
-i(
D^{AB}_I{}^i{}_jQ^C_I{}^j{}_\bi
-D^{AB}_{I+1}{}^\bj{}_\bi Q^C_I{}^i{}_\bj
)u_A^-u_B^-u_C^-
~.
\end{eqnarray}
The physical fields are $SU(2)$ doublet complex scalars
%(to be denoted as 
$Q^A$
%contained in $F^+$
and a pair of $SU(2)$ isosinglet spinors,
$\psi$ and $\kappa$.
%where $SU(2)_A$ is the automorphism of $\SN=2$

%%%%%%%%%%%%%%%%%%%%%%%%%%%%%%%%%%%%%%%%%%%%%%%%%%%%%%%%%%%%%%%%%
%%% references
%%%%%%%%%%%%%%%%%%%%%%%%%%%%%%%%%%%%%%%%%%%%%%%%%%%%%%%%%%%%%%%%%

\end{document}